\documentclass[aps,pre,
showkeys,
longbibliography,
reprint,                %% preprint, draft, 
twocolumn,              %% onecolumn, 
nofootinbib,
floatfix]{revtex4-2}

\usepackage[T1]{fontenc}
\usepackage[utf8]{inputenc}
\usepackage{amsmath}
\usepackage{amsfonts}
\usepackage{graphicx}
\usepackage[labelformat=simple]{subcaption}

\usepackage{xcolor}
\definecolor{codegreen}{rgb}{0,0.6,0}
\definecolor{codegray}{rgb}{0.5,0.5,0.5}
\definecolor{codepurple}{rgb}{0.58,0,0.82}
\definecolor{backcolour}{rgb}{0.95,0.95,0.92}

\usepackage{listings}
\lstdefinestyle{mystyle}{
    backgroundcolor=\color{backcolour},   
    commentstyle=\color{codegreen},
    keywordstyle=\color{magenta},
    numberstyle=\tiny\color{codegray},
    stringstyle=\color{codepurple},
    basicstyle=\ttfamily\footnotesize, %% tiny
    breakatwhitespace=false,         
    breaklines=true,                 
    captionpos=t,                    
    keepspaces=true,                 
    numbers=left, numberstyle=\tiny, stepnumber=2, numbersep=5pt,
    frame=lines,
    showspaces=false,                
    showstringspaces=true,
    breakautoindent=false,
    showtabs=false,                  
    tabsize=2
}
\usepackage{lipsum}

\usepackage[colorlinks=true,hyperfootnotes=true,breaklinks=true,allcolors=teal]{hyperref}
\usepackage[capitalise]{cleveref}

\begin{document}

\title{Bak--Tang--Wiesenfeld model for various topologies and ranges of interaction}

\author{Paweł Szczepaniak}
\affiliation{\href{https://ror.org/00bas1c41}{AGH University}, Faculty of Physics and Applied Computer Science, al.~Mickiewicza~30, 30-059 Krak\'ow, Poland}

\author{Krzysztof Malarz}
\email{malarz@agh.edu.pl}
\thanks{ORCID: \href{https://orcid.org/0000-0001-9980-0363}{0000-0001-9980-0363}}
\affiliation{\href{https://ror.org/00bas1c41}{AGH University}, Faculty of Physics and Applied Computer Science, al.~Mickiewicza~30, 30-059 Krak\'ow, Poland}

\begin{abstract}
In this paper, the Bak--Tang--Wiesenfeld model for various substrate topologies and a variety of neighborhoods is reconsidered. 
With computer simulation, we study the distribution of avalanche sizes.
Using the Z-score we confirm that independently of the substrate topology and the range of neighborhood, the exponent that governs the power law of the probability distribution of the size of avalanches is the same and approximately equal $1.208(39)$. 
However, this requires a smartly chosen number of deposited grains in relation to the linear size of the system.
\end{abstract}

\date{May 15th, 2026} %% \today


\maketitle

%% ############################################################
\section{Introduction}
%% ############################################################

Self-organizing criticality (SOC) \cite{Bak_1996} is a property of dynamical systems where once reached, the system persists indefinitely despite the occurrence of arbitrarily large (critical) fluctuations within it. Examples of such systems range from stock exchange markets \cite{Dmitriev_2024} (where fluctuations concern stock prices and critical fluctuations manifest as stock market crashes occurring over days), forming traffic jams \cite{Laval_2023}, appearing earthquakes \cite{Ida_2024}, cellular automata \cite{Antoniuk1998}, to the evolution of life on our planet \cite{Hewzulla_1999} (where fluctuations concern the number of species of living organisms and critical fluctuations manifest as mass extinctions occurring over hundreds of years). In all cases, the size and duration of the `catastrophe' obey a power law, meaning that the probability distribution of time and sizes has no characteristic scale: there are many small `catastrophes', but occasionally the size of the `catastrophe' approaches the size of the system. Examples of such a situation were Black Thursday on October 24, 1929 on Wall Street \cite{Carlson_2023} or the extinction of 98\% of marine and 70\% of land species during the End Permian mass extinction, which occurred around 255 million years ago \cite{Raup_1982}.

Much less spectacular effects are observed in the fluctuations of a sand pile, to which dynamics is introduced by successive trickling grains of sand. The angle of inclination of the pile with the ground then fluctuates, and `catastrophes' manifest themselves as avalanches of sand deposits along the side of the pile. Here, too, the duration and size $s$ of the avalanches are subject to a power law
%% ============================================================
\begin{equation}
\label{eq:Ps}
    P(s)\propto s^{-\tau}
\end{equation}
%% ============================================================
and constitute the archetype of systems that exhibit SOC, a point first highlighted in \citeauthor{PhysRevLett.59.381} work, where the exponent $\tau$ was estimated as close to one, as for the rouge noise \cite{PhysRevLett.59.381}.

The seminal paper of \citeauthor{PhysRevLett.59.381} started attempts to verify the hypothesis included in Reference~\onlinecite{PhysRevLett.59.381} in various ways and with various modifications \cite{PhysRevLett.134.187201,Poncelet_2017,Huynh_2011,Azimi-Tafreshi_2010,Koza_2007,Laurson_2005,Priezzhev_1997,Chung_1993,Dhar_1990}.
Among them, \citeauthor{Laurson_2005} proved that the exponent $\tau$ in \Cref{eq:Ps} is smaller than 2 \cite{Laurson_2005}.
The most recent paper on this topic \cite{PhysRevLett.134.187201} indicates $\tau\approx 1.26$.

The shape of the underlying lattice (square, triangular, honeycomb in two-dimensional space) may influence the critical system behavior by changing, for example, the critical temperature among interaction spins \cite{Malarz2005}, percolation thresholds \cite{2006.15621,2204.12593}, cross-over or phase transition temperature in structural balance \cite{2407.02603}.
The same happens for changing the range of interaction (and going further beyond the nearest neighbors) in percolation \cite{2006.15621,2204.12593}, opinion formation \cite{2405.05114}, thermodynamic and transport properties of a fluid \cite{Suvarna_2024}, topological systems \cite{Simonyan_2026}, etc.

In this paper, we check how various substrate topologies and ranges of interaction influence the probability distribution of the avalanche sizes.
Namely, we check shape of the above mentioned distribution: (i) for square, triangular, and honeycomb lattices; (ii) for neighborhoods containing only the nearest or the nearest and next-nearest neighbors; (iii) and various lattice sizes; (iv) and various number of deposited particles.

The results of Monte Carlo simulations, tested with the Z test, show that the exponents $\tau$ in the power law \eqref{eq:Ps} are statistically the same independently of the substrate topology and the assumed range of interactions. 

The paper is organized as follows: the next \Cref{sec:model} defines the model; \Cref{sec:results} presents the results of the computer simulation. 
The results are discussed and summarized in \Cref{sec:discussion}.
\Cref{app:Ps-NL} contains extra plots of dependencies $P(s)$ for various substrate sizes and the number of deposited grains, while \Cref{app:source} shows examples of Fortran implementations of extended neighborhoods with open boundary conditions.

%% ############################################################
\section{\label{sec:model}Model}
%% ############################################################

%% ============================================================
\begin{figure*}
%% ------------------------------------------------------------
\begin{subfigure}[t]{0.15\textwidth}
\caption{\label{fig:sq-1}}
\includegraphics{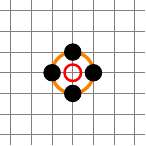}
\end{subfigure}
%% ------------------------------------------------------------
\hfill
%% ------------------------------------------------------------
\begin{subfigure}[t]{0.15\textwidth}
\caption{\label{fig:sq-2}}
\includegraphics{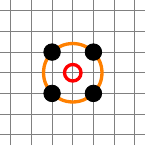}
\end{subfigure}
%% ------------------------------------------------------------
\hfill
%% ------------------------------------------------------------
\begin{subfigure}[t]{0.15\textwidth}
\caption{\label{fig:tr-1}}
\includegraphics{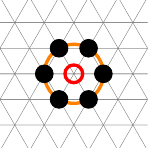}
\end{subfigure}
%% ------------------------------------------------------------
\hfill
%% ------------------------------------------------------------
\begin{subfigure}[t]{0.15\textwidth}
\caption{\label{fig:tr-2}}
\includegraphics{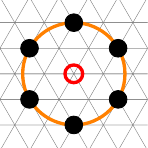}
\end{subfigure}
%% ------------------------------------------------------------
\hfill
%% ------------------------------------------------------------
\begin{subfigure}[t]{0.15\textwidth}
\caption{\label{fig:hc-1}}
\includegraphics{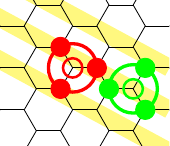}
\end{subfigure}
%% ------------------------------------------------------------
\hfill
%% ------------------------------------------------------------
\begin{subfigure}[t]{0.15\textwidth}
\caption{\label{fig:hc-2}}
\includegraphics{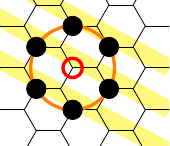}
\end{subfigure}
%% ------------------------------------------------------------
\caption{\label{fig:lattice}\subref{fig:sq-1}--\subref{fig:sq-2} Square, \subref{fig:tr-1}--\subref{fig:tr-2} triangular and \subref{fig:hc-1}--\subref{fig:hc-2} honeycomb lattices with marked neighborhoods in the first \subref{fig:sq-1}, \subref{fig:tr-1}, \subref{fig:hc-1} and the first plus the second \subref{fig:sq-2}, \subref{fig:tr-2}, \subref{fig:hc-2}  coordination zone. As honeycomb lattice does not belong to the Bravais' lattices, the sites in the first coordination zone are different in odd and even rows and columns}
\end{figure*}
%% ============================================================

%% ============================================================
\begin{figure*}
%% ------------------------------------------------------------
\begin{subfigure}[t]{0.14\textwidth}
\caption{\label{fig:comp-sq-1}}
\includegraphics{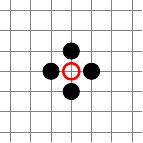}
\end{subfigure}
%% ------------------------------------------------------------
\hfill
%% ------------------------------------------------------------
\begin{subfigure}[t]{0.14\textwidth}
\caption{\label{fig:comp-sq-12}}
\includegraphics{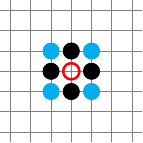}
\end{subfigure}
%% ------------------------------------------------------------
\hfill
%% ------------------------------------------------------------
\begin{subfigure}[t]{0.14\textwidth}
\caption{\label{fig:comp-tr-1}}
\includegraphics{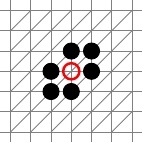}
\end{subfigure}
%% ------------------------------------------------------------
\hfill
%% ------------------------------------------------------------
\begin{subfigure}[t]{0.14\textwidth}
\caption{\label{fig:comp-tr-12}}
\includegraphics{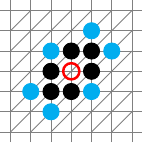}
\end{subfigure}
%% ------------------------------------------------------------
\hfill
%% ------------------------------------------------------------
\begin{subfigure}[t]{0.16\textwidth}
\caption{\label{fig:comp-hc-1}}
\includegraphics{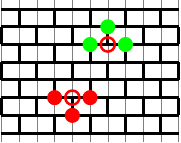}
\end{subfigure}
%% ------------------------------------------------------------
\hfill
%% ------------------------------------------------------------
\begin{subfigure}[t]{0.16\textwidth}
\caption{\label{fig:comp-hc-12}}
\includegraphics{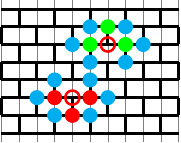}
\end{subfigure}
%% ------------------------------------------------------------
\caption{\label{fig:lattice-comput}Computerized version of lattices shown in \Cref{fig:lattice} mapped into square lattice:
\subref{fig:comp-sq-1} \textsc{sq-1}; 
\subref{fig:comp-sq-12} \textsc{sq-1,2}; 
\subref{fig:comp-tr-1} \textsc{tr-1};
\subref{fig:comp-tr-12} \textsc{tr-1,2};
\subref{fig:comp-hc-1} \textsc{hc-1};
\subref{fig:comp-hc-12} \textsc{hc-1,2}
}
\end{figure*}
%% ============================================================

The computer model of the sand pile is defined on three different types of substrate and two different types of neighborhood presented in \Cref{fig:lattice}.
For neighborhood names, we keep the convention proposed in Reference~\onlinecite{2010.02895}.
Alphanumerical strings encode both: the lattice shape (\textsc{sq}, \textsc{tr}, \textsc{hc} for square, triangular, and honeycomb, respectively) and the range of interaction (1 and 1,2 for the nearest neighbors, and the nearest plus next-nearest neighbors, respectively).
Initially, at time $t=0$, the substrate is flat.
In each time step $t>0$, a single position $i$ is randomly chosen and a single grain is added to the top of the selected site.
If the height $h(i)$ of the column at the site $i$ is greater than or equal to the predefined value of $K$, then the particles of this column fall to $K$ their neighbors: $h(i)\to h(i)-K$ and simultaneously $h(n)\to h(n)+1$ for every site $n$ that is in the neighborhood of the site $i$.
The values of $K$ for various lattices and neighborhoods are presented in \Cref{tab:tau}.
The grains falling to the neighboring sites $j$ can initiate yet another condition $h(j)\ge K$ and produce recursive falling grains from the site $j$ to its neighbors, etc.
This recursive procedure lasts until all sites $i$ in the lattice meet the condition $h(i)<K$.
Then the time $t$ is incremented and the next site is selected to accept the new grain.
The number of particles moved from site to site between the time steps $t$ and $t+1$ defines the size $s$ of the observed avalanche.
If the grain reaches the lattice border, its move is accounted for the size of the avalanche, but it leaves the system.

The lattices and their neighborhoods presented in \Cref{fig:lattice} are encoded as a one-dimensional vector of length $L^2$, where $L$ is the linear size of the square lattice, which is also the backbone for the triangular and honeycomb lattices.
The computer-useful presentation of the mapping lattices and neighborhoods considered here is presented schematically in \Cref{fig:lattice-comput}.
Then, the site $i$ has four neighbors ($K=4$) at $i\pm 1$ and $i\pm L$ for the square lattice and the nearest-neighbors (see \Cref{fig:comp-sq-1}) or six neighbors ($K=6$) at $i\pm 1$, $i\pm L$, $i+L+1$, $i-L-1$ for the triangular lattice and the nearest-neighbors (see \Cref{fig:comp-tr-1}).
The honeycomb lattice can be converted to a brick-wall lattice representation (see \Cref{fig:comp-hc-1,fig:comp-hc-12}).
Unfortunately, as honeycomb lattice is excluded from Bravais lattices, their one-dimensional representation must be split to odd and even numbers of rows and columns. 
An example of Fortran implementation of \textsc{hc-1,2} neighborhood is presented in \Cref{lst:hc-12} in \Cref{app:source}.
Additionally, \Cref{lst:tr-12,lst:sq-12} in \Cref{app:source} show more obvious implementations of \textsc{sq-1,2} and \textsc{tr-1,2} neighborhoods (also as Fortran subroutines), respectively.

%% ############################################################
\section{\label{sec:results}Results}
%% ############################################################

\begin{figure}[btp]
%% \psfrag{lg(s)}{$\lg(s)$}
%% \psfrag{lg(P(s))}{$\lg P(s)$}
%% \psfrag{sq-1}[][r]{\textsc{sq-1}}
%% \psfrag{hc-1}[][r]{\textsc{hc-1}}
%% \psfrag{tr-1}[][r]{\textsc{tr-1}}
%% \psfrag{sq-12}[][r]{\textsc{sq-1,2}}
%% \psfrag{hc-12}[][r]{\textsc{hc-1,2}}
%% \psfrag{tr-12}[][r]{\textsc{tr-1,2}}
%% \includegraphics[width=\columnwidth]{soc.eps}
\includegraphics[width=\columnwidth]{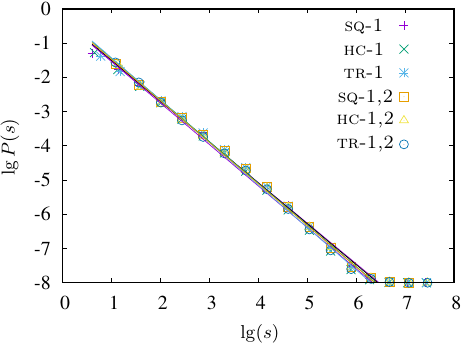}
\caption{\label{fig:soc}Probability distribution $P(s)$ of avalanches size $s$ logarithmically binned for $L=10^3$ and $N=10^8$ (for raw data look at orange symbols in \Cref{fig:Ps-N} presented in \Cref{app:Ps-NL})}
\end{figure}

In \Cref{fig:soc} the probability distribution functions $P(s)$ for observing the avalanche of size $s$  (for the linear size $L=10^3$ of the substrate and after deposition of $N=10^8$ particles) for various substrates and various ranges of the neighborhood are presented.
The data are logarithmically binned and the linear fit according to the least-square method is also plotted.
The raw data (before applying the binning procedure) are presented as orange symbols in \Cref{fig:Ps-N} in \Cref{app:Ps-NL}.

%% ============================================================
%% ============================================================

The results of the straight line fitting on logarithmic scale to $\lg P(s)$ vs. $\lg s$ correspond to the fitting power law \eqref{eq:Ps}.
The slopes obtained (corresponding to the exponent $\tau$) together with their uncertainties $u(\tau)$ are collected in \Cref{tab:tau}.

%% ============================================================
\begin{table}[tb]
\caption{\label{tab:tau}Exponent $\tau$ and its uncertainty $u(\tau)$ for various lattices and neighborhoods and linear lattice size $L=10^3$ after deposition of $N=10^8$ particles.
The data correspond to the least-square method fits presented in \Cref{fig:soc}}
\begin{ruledtabular}
\begin{tabular}{lrrrrrr}
       & \textsc{sq-1}   & \textsc{sq-1,2} & \textsc{tr-1} & \textsc{tr-1,2} & \textsc{hc-1} & \textsc{hc-1,2} \\ \hline
$K$       & 4            & 8               & 6             & 12              & 3             & 9    \\ 
$\tau$    & 1.190        & 1.214           & 1.196         & 1.230           & 1.209         & 1.210\\
$u(\tau)$ & 0.021        & 0.015           & 0.018         & 0.013           & 0.015         & 0.012\\         
\end{tabular}
\end{ruledtabular}
\end{table}
%% ============================================================

%% ============================================================
\begin{table}[tb]
\caption{\label{tab:Z}The pairwise absolute normalized difference [$Z$-score, \Cref{eq:z_test_eq}] values $|Z|$ among various lattices and neighborhoods. Values of $\tau$ and their uncertainness $u(\tau)$ are taken after \Cref{tab:tau}}
\begin{ruledtabular}
\begin{tabular}{lrrrrrr}
 & \textsc{hc-1,2} &	\textsc{hc-1} &	\textsc{tr-1,2}	& \textsc{tr-1}	& \textsc{sq-1,2} &	\textsc{sq-1} \\ \hline
%%               hc-1,2 & hc-1	& tr-1,2& tr-1 & sq-1,2& sq-1
\textsc{sq-1}	& 0.83	& 0.74	& 1.62	& 0.22 & 0.93  & 0.00 \\
\textsc{sq-1,2}	& 0.21	& 0.24	& 0.81	& 0.77 & 0.00  &	  \\
\textsc{tr-1}	& 0.65	& 0.55	& 1.53	& 0.00 &	   &	  \\
\textsc{tr-1,2}	& 1.13	& 1.06	& 0.00	&      &	   &	  \\
\textsc{hc-1}	& 0.05	& 0.00	&       &      &	   &	  \\
\textsc{hc-1,2}	& 0.00	&		&		&      &	   &	  \\
\end{tabular}
\end{ruledtabular}
\end{table}
%% ============================================================

We set the null hypothesis that the values obtained for $\tau$ are statistically the same within the estimated uncertainties $u(\tau)$.
To test the null hypothesis, we compute the normalized difference
\begin{equation}
Z=\frac{\tau_1-\tau_2 }{\sqrt{u(\tau_1)^2+u(\tau_2)^2}},
\label{eq:z_test_eq}
\end{equation}
where indexes 1 and 2 correspond to pairs of selected lattices and neighborhoods.
The absolute normalized difference obtained pairwise for the values of $\tau$ and their uncertainties given in \Cref{tab:tau} are presented in \Cref{tab:Z}.

%% ############################################################
\section{\label{sec:discussion}Discussion}
%% ############################################################

In this paper, the BTW model is reconsidered.
We check how various substrate topologies and the assumed ranges of interactions influence the probability distribution $P(s)$ for the size $s$ of the avalanches.
As one may expect, for a fixed number $L^2$ of available sites for particle deposition, the increase of deposited particles $N$ shifts the tail of the distribution $P(s)$ towards higher values of $s$ at lower values $P(s)$ removing, subsequently, fat-tails of the distribution $P(s)$ (see \Cref{fig:Ps-N} in \Cref{app:Ps-NL}).
For the largest values of deposited particles $N=10^8$ and the lattice size $L=10^3$ the distribution $P(s)$ follows nicely the power law \eqref{eq:Ps} with exponent $\tau<2$ as predicted by \citeauthor{Laurson_2005} in Reference \onlinecite{Laurson_2005}.
The mean value of $\bar\tau$ indicated in \Cref{tab:tau} is 1.208(39), which is consistent with the value 1.26 very recently reported in Reference~\onlinecite{PhysRevLett.134.187201}.

However, we notice that the possibility of observation of the power law \eqref{eq:Ps} is strongly influenced by the factor of particles deposited relative to the number of available sites $N/L^2$ (see \Cref{fig:Ps-N,fig:Ps-L} in \Cref{app:Ps-NL}).

The highest absolute normalized difference value $|Z|=1.62$ is observed for pair \textsc{sq-1} and \textsc{tr-1,2}.
The corresponding two-sided $p$-value is approximately $p\approx 0.13$.
So, the two measurements of $\tau$ are statistically consistent with each other at the usual 95\% confidence level.
The same occurs for all pairs lattice/neighborhood discussed here.

To conclude, for a reasonably selected factor $N/L^2$, the obtained exponent values $\tau$ are statistically the same, independently of the substrate structure and the assumed interaction range (at the 95\% confidence level).

Further studies may focus on applying the BTW model to other substrate topologies or richer neighborhoods or even simulation of avalanche propagation in higher (nonphysical) dimensions.
Furthermore, since \Cref{fig:Ps-N,fig:Ps-L} show the dependence of the value of the exponent $\tau$ on the number of grains ($N$) and the linear size of the system ($L$), to clarify the issue of universality of the exponent $\tau$, a systematic analysis of the finite size-effects in the parameters $N$ and $L$ may be performed.

\begin{acknowledgments}
This research was supported by a subsidy from the Polish Ministry of Science and Higher Education.
\end{acknowledgments}

%% \clearpage

\appendix

%% ############################################################
\section{\label{app:Ps-NL}Dependencies $P(s)$ for various $N$ and $L$}
%% ############################################################

\begin{figure*}
%% ------------------------------------------------------------
%% \psfrag{s}{$s$}
%% \psfrag{P(s)}{$P(s)$}
%% \psfrag{N}{$N$}
%% ------------------------------------------------------------
\begin{subfigure}[t]{0.49\textwidth}
\caption{\label{fig:soc-sq-1}}
\includegraphics[width=\textwidth]{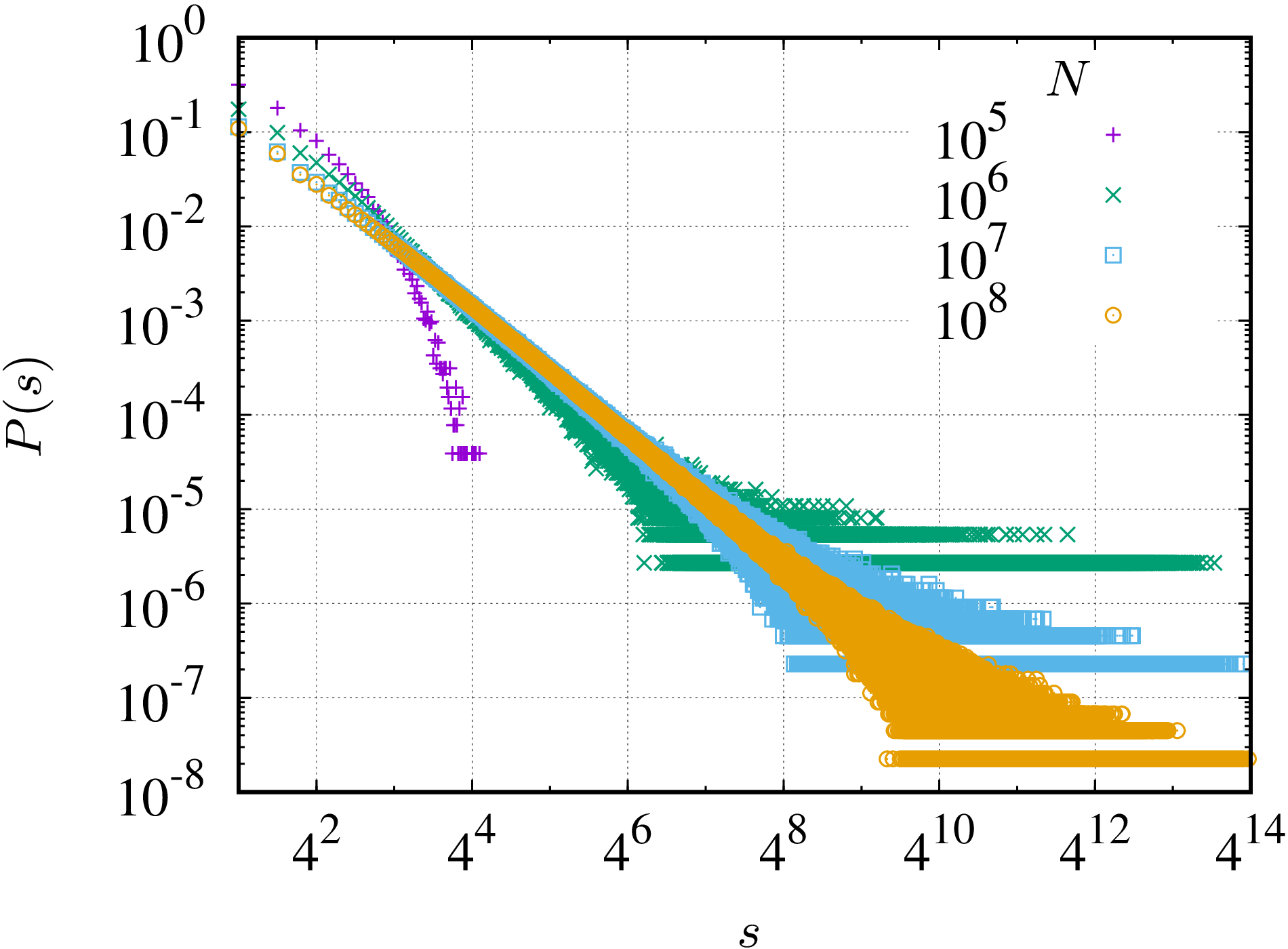}
\end{subfigure}
%% ------------------------------------------------------------
\begin{subfigure}[t]{0.49\textwidth}
\caption{\label{fig:soc-sq-12}}
\includegraphics[width=\textwidth]{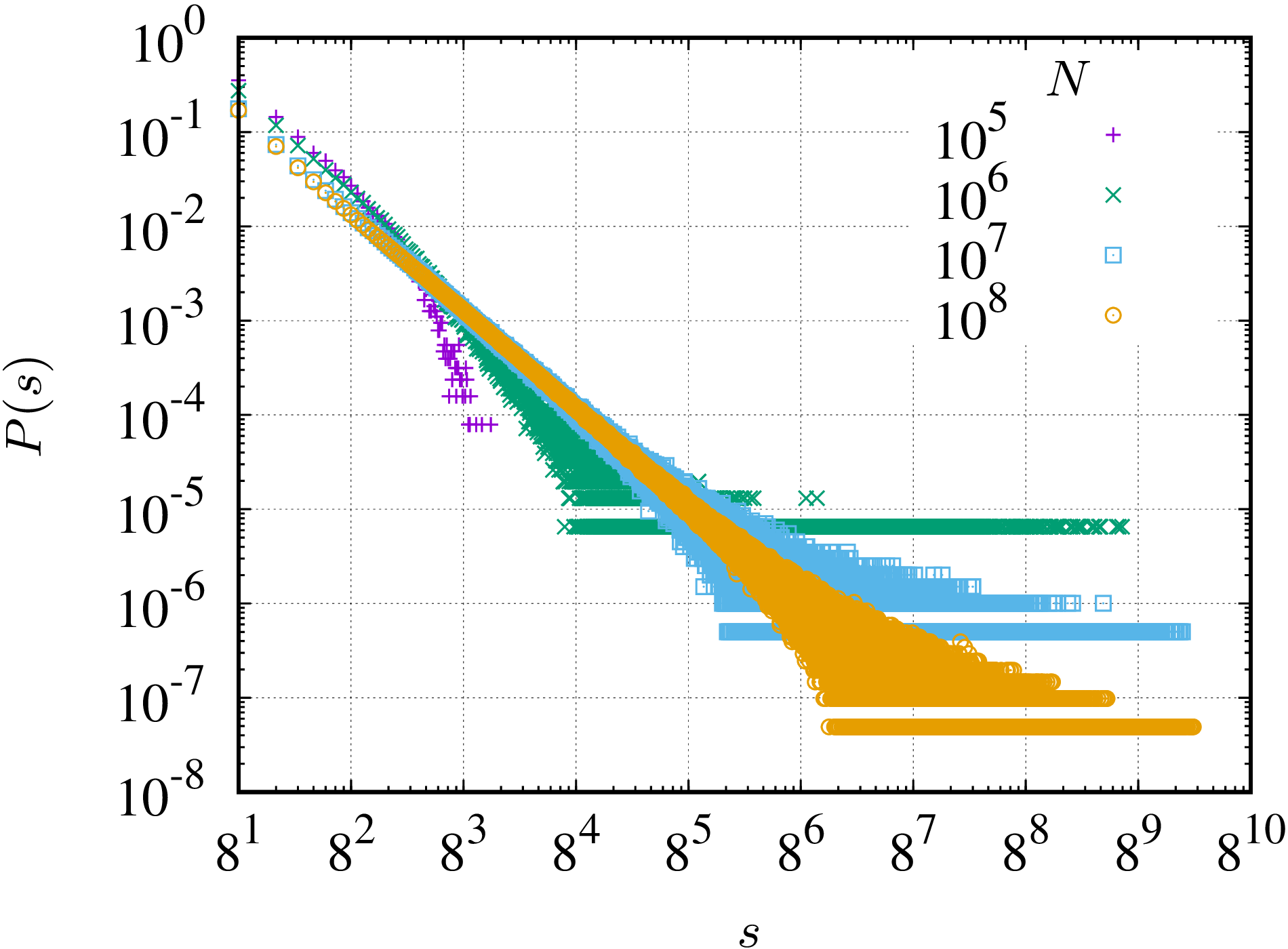}
\end{subfigure}
%% ------------------------------------------------------------
\begin{subfigure}[t]{0.49\textwidth}
\caption{\label{fig:soc-tr-1}}
\includegraphics[width=\textwidth]{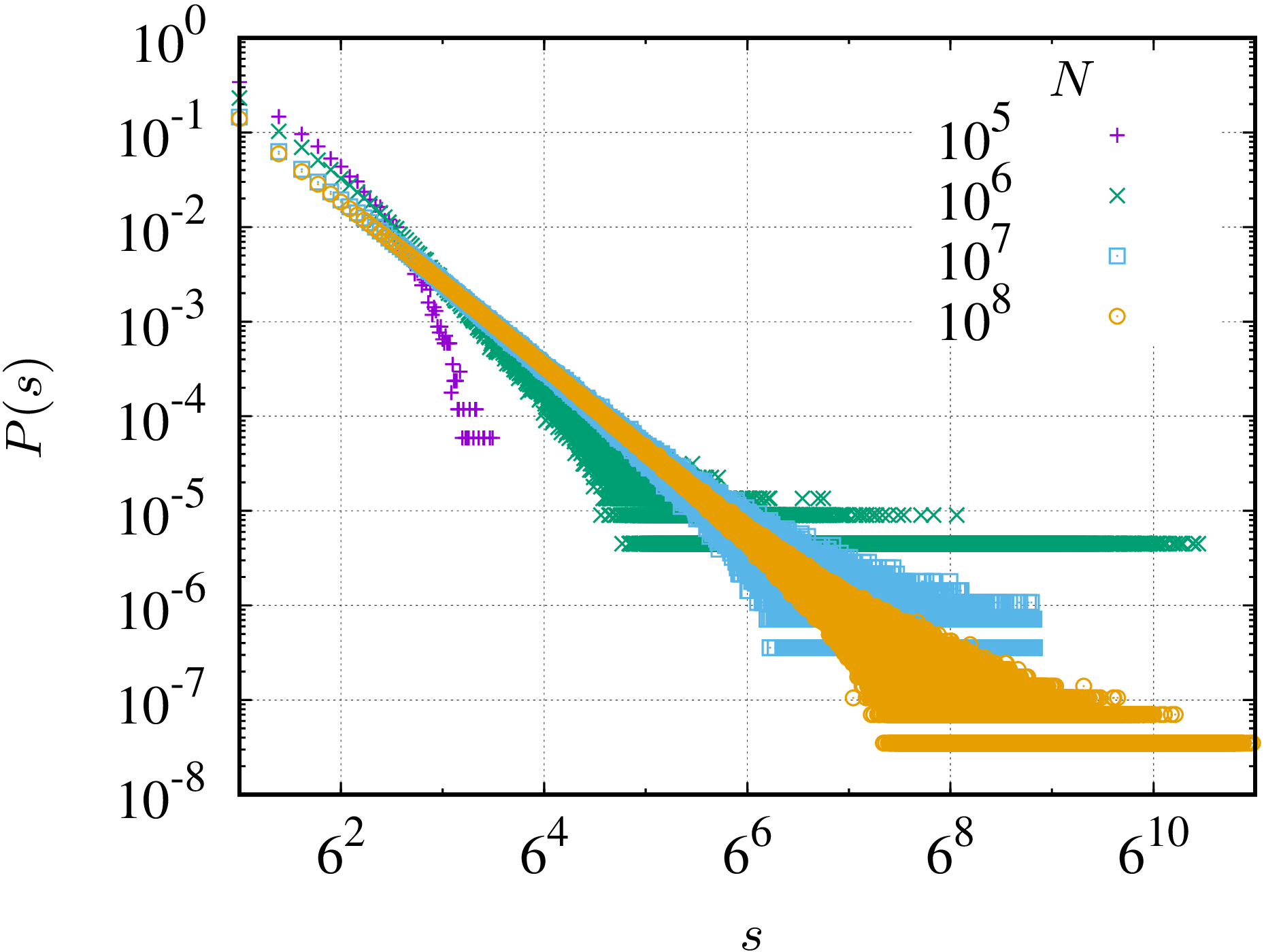}
\end{subfigure}
%% ------------------------------------------------------------
\begin{subfigure}[t]{0.49\textwidth}
\caption{\label{fig:soc-tr-12}}
\includegraphics[width=\textwidth]{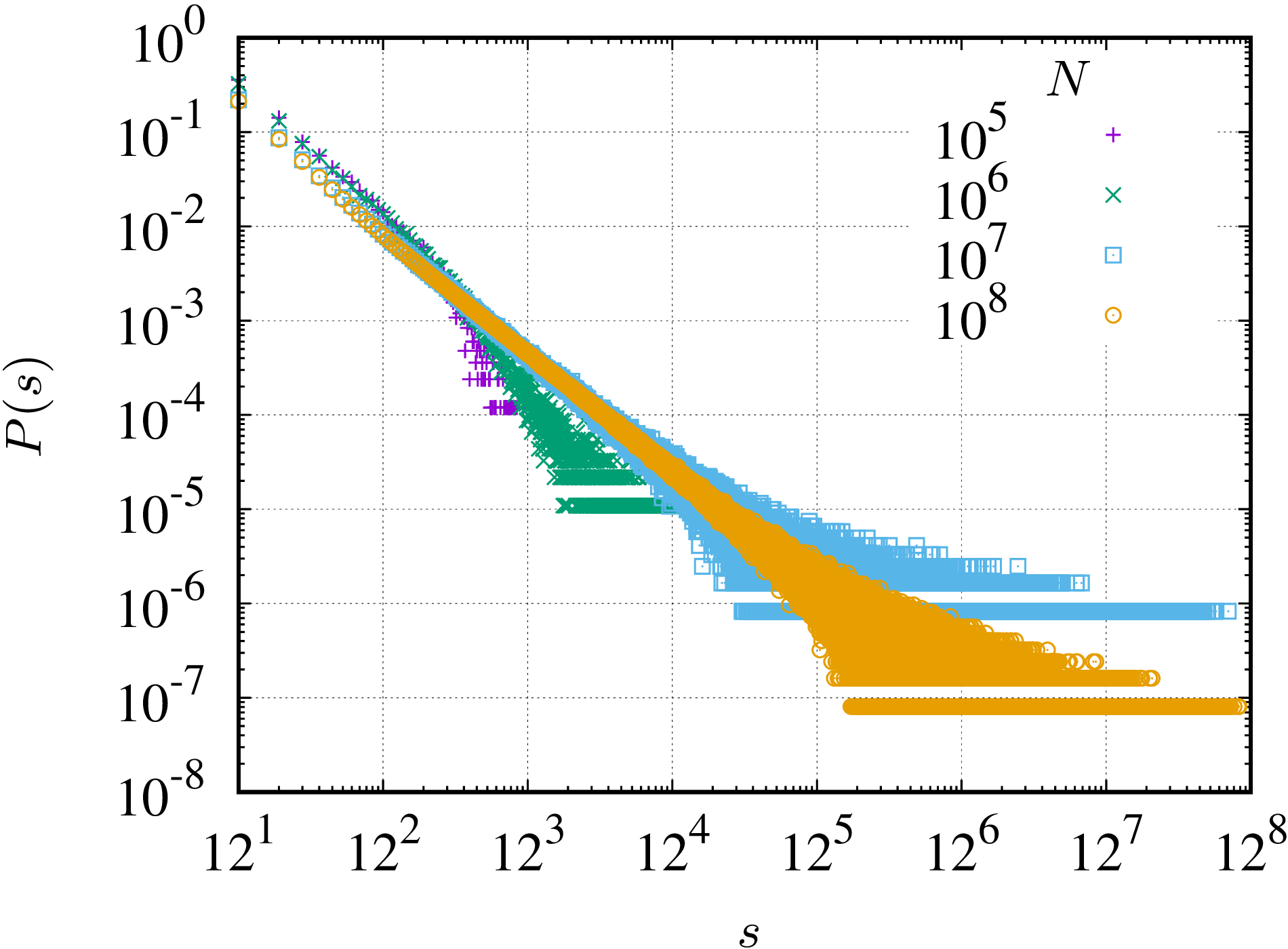}
\end{subfigure}
%% ------------------------------------------------------------
\begin{subfigure}[t]{0.49\textwidth}
\caption{\label{fig:soc-hc-1}}
\includegraphics[width=\textwidth]{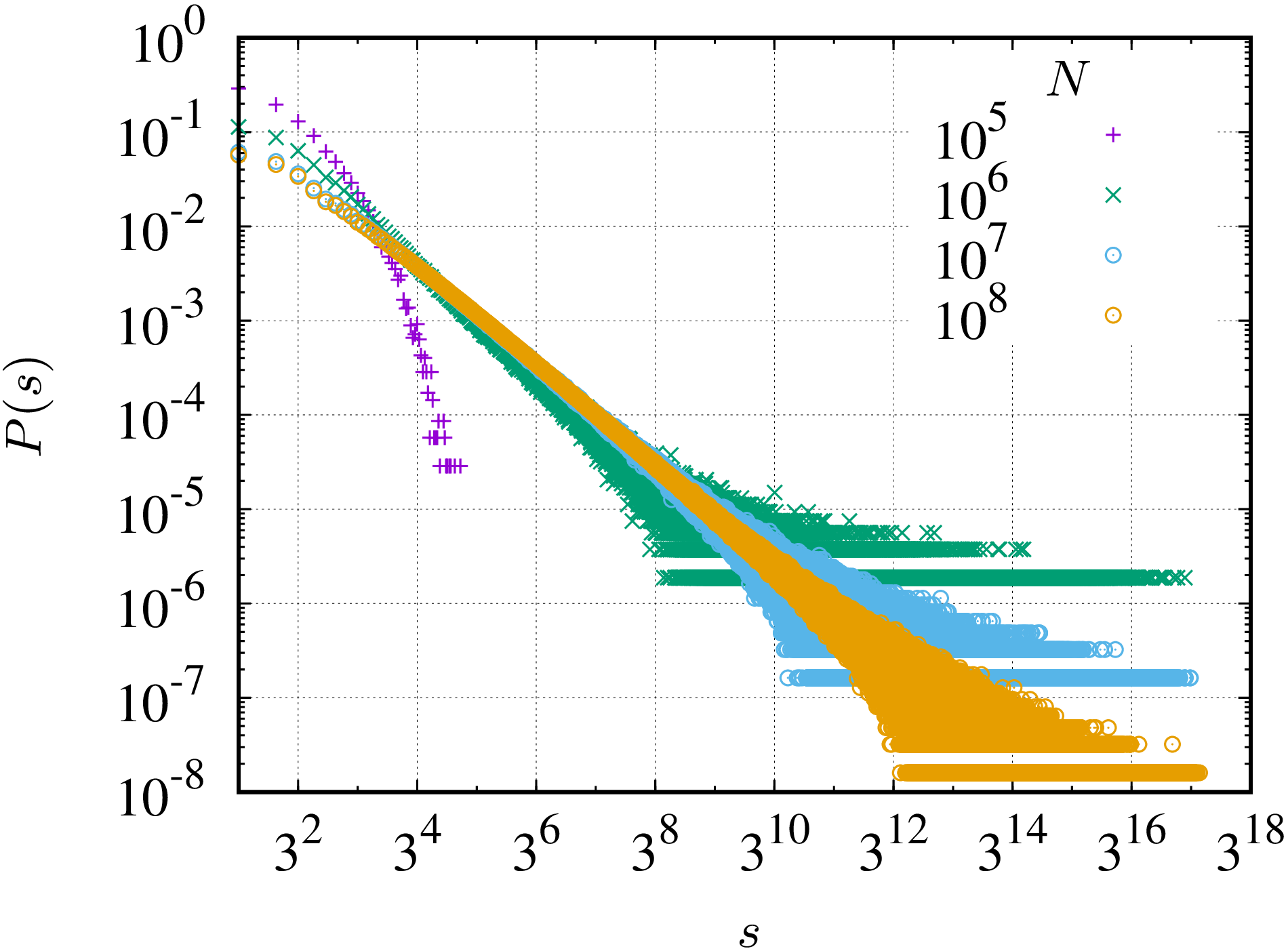}
\end{subfigure}
%% ------------------------------------------------------------
\begin{subfigure}[t]{0.49\textwidth}
\caption{\label{fig:soc-hc-12}}
\includegraphics[width=\textwidth]{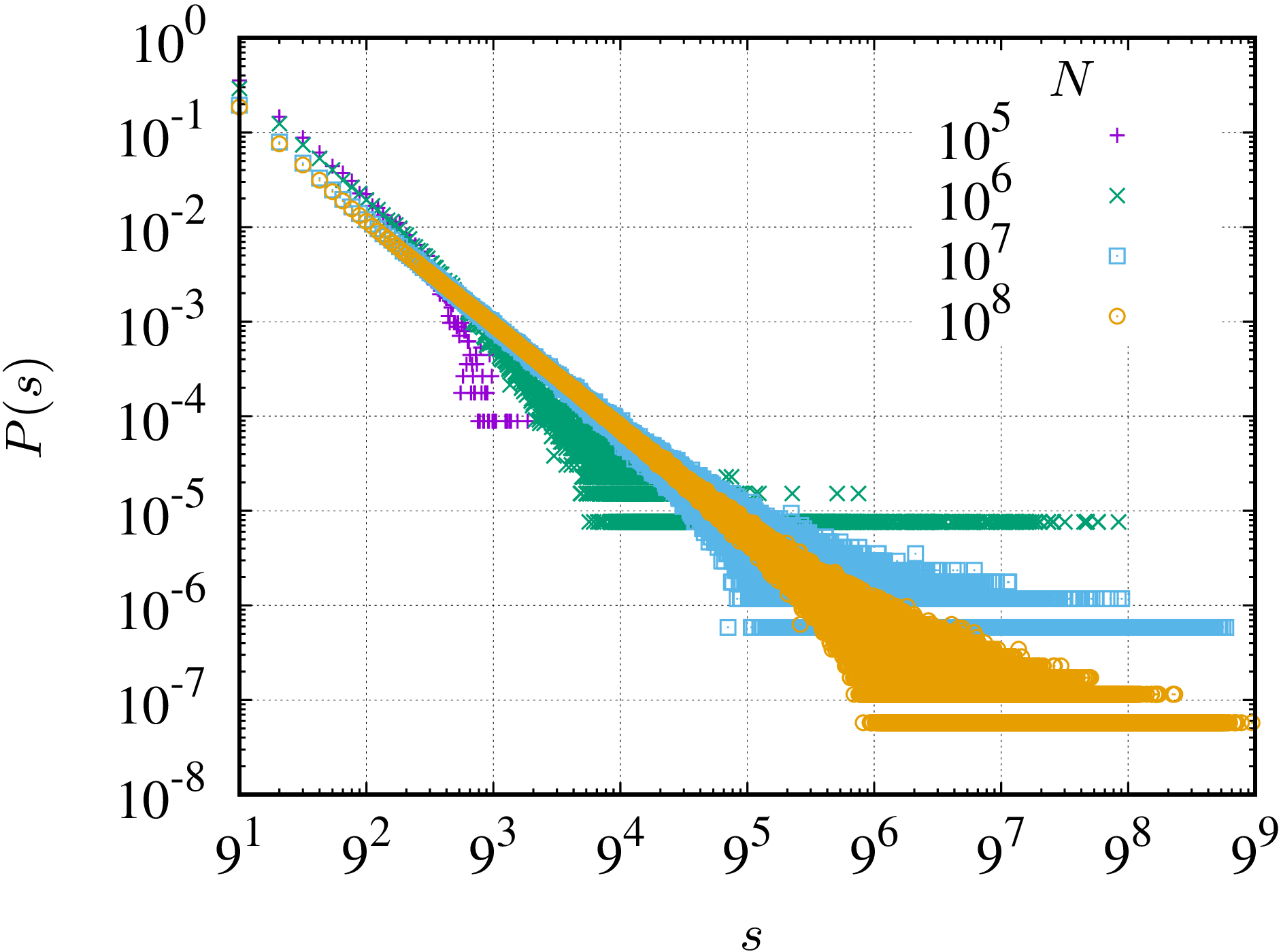}
\end{subfigure}
%% ------------------------------------------------------------
\caption{\label{fig:Ps-N}Probability distribution $P(s)$ of sizes of avalanches $s$ for various lattices and neighborhoods: 
\subref{fig:soc-sq-1} \textsc{sq-1}, 
\subref{fig:soc-sq-12} \textsc{sq-1,2}, 
\subref{fig:soc-tr-1} \textsc{tr-1}, 
\subref{fig:soc-tr-12} \textsc{tr-1,2}, 
\subref{fig:soc-hc-1} \textsc{hc-1}, 
\subref{fig:soc-hc-12} \textsc{hc-1,2}. 
Various number $N$ of deposited particles, fixed number of sites $L^2=10^6$}
\end{figure*}

\Cref{fig:Ps-N} shows the probability distributions $P(s)$ of observing avalanches of size $s$ for the fixed number of sites $L^2=10^6$ and the various numbers $N$ of deposited grains.
The plots shown in \Cref{fig:Ps-N} correspond to various lattices and assumed neighborhoods.

The insufficient number of deposited particles ($N=10^5$) does not allow one to observe the power law \eqref{eq:Ps}.
In this case, the sites with high columns $h(i)=K-1$ are separated by too long distances, and avalanches involving a lot of grains are rather rare.
The latter results in the drop of the $P(s)$ curve below the straight line predicted by the SOC theory.

\begin{figure*}
%% ------------------------------------------------------------
%% \psfrag{s}{$s$}
%% \psfrag{P(s)}{$P(s)$}
%% \psfrag{L}{$L$}
%% ------------------------------------------------------------
\begin{subfigure}[t]{0.49\textwidth}
\caption{\label{fig:soc-sq-1-L}}
\includegraphics[width=\textwidth]{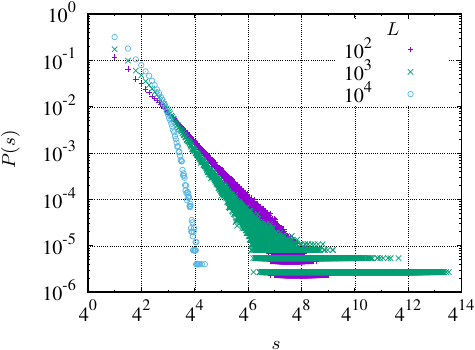}
\end{subfigure}
%% ------------------------------------------------------------
\begin{subfigure}[t]{0.49\textwidth}
\caption{\label{fig:soc-sq-12-L}}
\includegraphics[width=\textwidth]{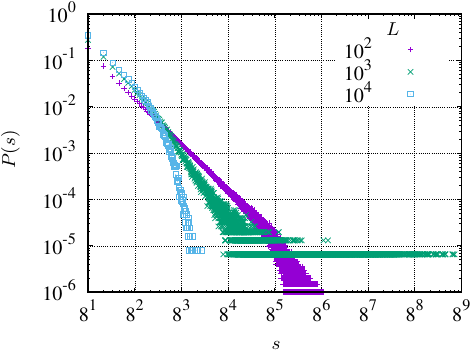}
\end{subfigure}
%% ------------------------------------------------------------
\begin{subfigure}[t]{0.49\textwidth}
\caption{\label{fig:soc-tr-1-L}}
\includegraphics[width=\textwidth]{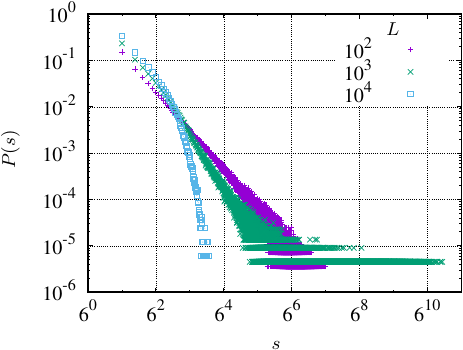}
\end{subfigure}
%% ------------------------------------------------------------
\begin{subfigure}[t]{0.49\textwidth}
\caption{\label{fig:soc-tr-12-L}}
\includegraphics[width=\textwidth]{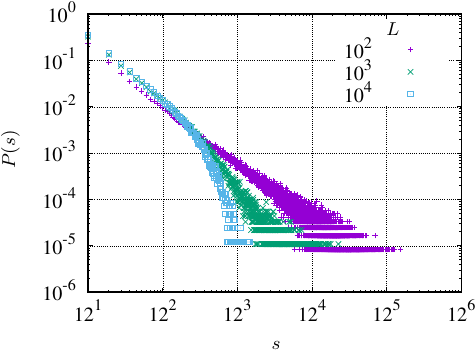}
\end{subfigure}
%% ------------------------------------------------------------
\begin{subfigure}[t]{0.49\textwidth}
\caption{\label{fig:soc-hc-1-L}}
\includegraphics[width=\textwidth]{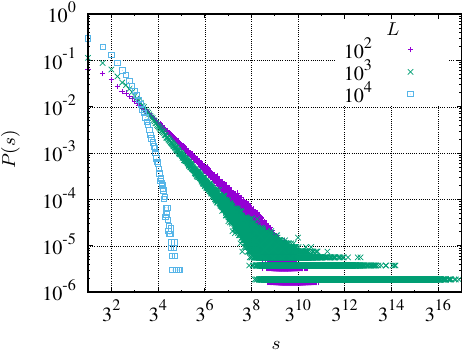}
\end{subfigure}
%% ------------------------------------------------------------
\begin{subfigure}[t]{0.49\textwidth}
\caption{\label{fig:soc-hc-12-L}}
\includegraphics[width=\textwidth]{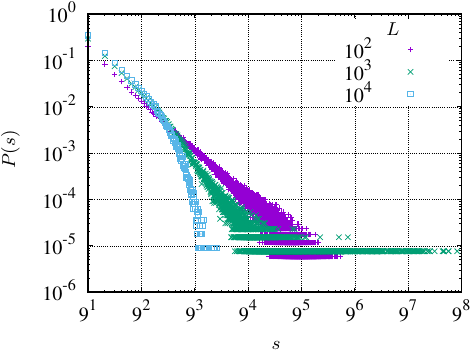}
\end{subfigure}
%% ------------------------------------------------------------
\caption{\label{fig:Ps-L}Probability distribution $P(s)$ of sizes of avalanches $s$ for various lattices and neighborhoods: 
\subref{fig:soc-sq-1-L} \textsc{sq-1}, 
\subref{fig:soc-sq-12-L} \textsc{sq-1,2}, 
\subref{fig:soc-tr-1-L} \textsc{tr-1}, 
\subref{fig:soc-tr-12-L} \textsc{tr-1,2},
\subref{fig:soc-hc-1-L} \textsc{hc-1}, 
\subref{fig:soc-hc-12-L} \textsc{hc-1,2}. 
Various number of sites $L^2$ for fixed number of deposited particles $N=10^6$}
\end{figure*}

\Cref{fig:Ps-L} shows the probability distributions $P(s)$ of observing avalanches of size $s$ for the fixed number of the deposited number $N=10^6$ of grains and the various number $L^2$ of available sites.
The plots shown in \Cref{fig:Ps-L} correspond to various lattices and assumed neighborhoods.

Also, here, the deviation from \Cref{eq:Ps} is particularly spectacular, when the system size ($L^2=10^4$) is too large compared to the number ($N=10^6$) of particles deposited.

%% ============================================================
%% ============================================================

%% ============================================================
%% ============================================================

%% ############################################################
\section{\label{app:source}Examples of boundaries subroutines}
%% ############################################################

\Cref{lst:sq-12,lst:tr-12,lst:hc-12} show Fortran subroutines {\tt boundaries()} implementing neighbors (stored in matrix $nn(i,k)$) of site $i$ for the nearest (1NN) and the next-nearest (2NN) neighbors for square, triangular, and honeycomb lattices, respectively.
The index $i=Lr+c$ and $L$ is the linear size of the underlying square lattice, and $r$ and $c$ enumerate the rows and columns of this lattice, respectively.
The index $1\le k\le K$ enumerates neighbors.

\begin{lstlisting}[language=Fortran,caption={\tt boundaries()} function for the \textsc{sq-1,2} neighborhood,label=lst:sq-12]
      SUBROUTINE boundaries()
      IMPLICIT none
      INTEGER K,h,nn,L,L2,i,c,r
      PARAMETER (K=8,L=1e3,L2=L*L)
      DIMENSION h(0:L2-1),nn(0:L2-1,K)
      COMMON h,nn

      DO c=1,L-2
      DO r=1,L-2
        i=L*r+c
C 1NN: ----------------------------------------
        nn(i,1) = i -1
        nn(i,2) = i +1
        nn(i,3) = i -L
        nn(i,4) = i +L
C 2NN: ----------------------------------------
        nn(i,5) = i -L-1
        nn(i,6) = i -L+1
        nn(i,7) = i +L-1
        nn(i,8) = i +L+1
      ENDDO
      ENDDO
      END SUBROUTINE boundaries
\end{lstlisting}

\begin{lstlisting}[language=Fortran,caption={\tt boundaries()} function for the \textsc{tr-1,2} neighborhood,label=lst:tr-12]
      SUBROUTINE boundaries()
      IMPLICIT none
      INTEGER K,h,nn,L,L2,i,c,r
      PARAMETER (K=12,L=10,L2=L*L)
      DIMENSION h(0:L2-1),nn(0:L2-1,K)
      COMMON h,nn

      DO c=2,L-3
      DO r=2,L-3
        i=L*r+c
C 1NN: ----------------------------------------
        nn(i, 1) = i -1
        nn(i, 2) = i +1
        nn(i, 3) = i -L
        nn(i, 4) = i +L
        nn(i, 5) = i -L-1
        nn(i, 6) = i +L+1
C 2NN: ----------------------------------------
        nn(i, 7) = i +L-1
        nn(i, 8) = i -L+1
        nn(i, 9) = i +2*L+1
        nn(i,10) = i -2*L-1
        nn(i,11) = i -L-2
        nn(i,12) = i +L+2
      ENDDO
      ENDDO
      END SUBROUTINE boundaries
\end{lstlisting}

\begin{lstlisting}[language=Fortran,caption={\tt boundaries()} function for the \textsc{hc-1,2} neighborhood,label=lst:hc-12]
      SUBROUTINE boundaries()
      IMPLICIT none
      INTEGER K,h,nn,L,L2,i,r,c
      PARAMETER (K=9,L=1e3,L2=L*L)
      DIMENSION h(0:L2-1),nn(0:L2-1,K)
      COMMON h,nn

      DO c=2,L-3
      DO r=1,L-2
      i=L*r+c
C 1NN: ----------------------------------------
        nn(i,1) = i-1
        nn(i,2) = i+1
        IF(MOD(r,2).EQ.0) THEN
          IF(MOD(c,2).EQ.0) THEN
            nn(i,3) = i+L
          ELSE
            nn(i,3) = i-L
          ENDIF
        ELSE
          IF(MOD(c,2).EQ.0) THEN
            nn(i,3) = i-L
          ELSE
            nn(i,3) = i+L
          ENDIF
        ENDIF
C 2NN: ----------------------------------------
        nn(i,4) = i -2  
        nn(i,5) = i +2  
        nn(i,6) = i +L-1
        nn(i,7) = i +L+1
        nn(i,8) = i -L-1
        nn(i,9) = i -L+1
      ENDDO
      ENDDO
      END SUBROUTINE boundaries
\end{lstlisting}

%% ############################################################
%% \bibliography{basics,soc,percolation,km, this}
%% ############################################################

\end{document}